\documentclass[11pt]{revtex4}
\usepackage{amsmath}
\usepackage{amssymb}
\usepackage{amsfonts}
\usepackage{mathrsfs}
\usepackage{epsfig}
\usepackage{bm}
\begin{document}
\title{Spatial-temporal evolution of the current filamentation instability}

\author{V. B. Pathak$^{1,2}$\footnote{vishwa.bandhu@ist.utl.pt}, T. Grismayer $^1$, A. Stockem $^1$, R. A. Fonseca$^{1,3}$, L. O. Silva$^1$\footnote{luis.silva@ist.utl.pt} }
\address{$^1$GoLP/Instituto de Plasmas e Fus\~{a}o Nuclear, Instituto Superior T\'{e}cnico, Universidade de Lisboa, 1049-001 Lisboa, Portugal}
\address{$^2$ Center for Relativistic Laser Science, Institute for Basic Science (IBS), Gwangju 500-712, Korea}
\address{$^3$DCTI/ISCTE Lisbon University Institute, 1649-026 Lisbon, Portugal}

%----------------------
%   ABSTRACT
%----------------------

\begin{abstract}
The spatial-temporal evolution of the purely transverse current filamentation instability is analyzed by deriving a single partial differential equation for the instability and obtaining the analytical solutions for the spatially and temporally growing current filament mode. When the beam front always encounters fresh plasma, our analysis shows that the instability grows spatially from the beam front to the back up to a certain critical beam length; then the instability acquires a purely temporal growth. This critical beam length increases linearly with time and in the non-relativistic regime it is proportional to the beam velocity. In the relativistic regime the critical length is inversely proportional to the cube of the beam Lorentz factor $\gamma_{0b}$. Thus, in  the ultra-relativistic regime the instability immediately acquires a purely temporal growth all over the beam. The analytical results are in good agreement with multidimensional particle-in-cell simulations performed with OSIRIS. Relevance of current study to recent and future experiments on fireball beams is also addressed.

\end{abstract}

\maketitle
%---------------------------------
%   SEC.1: Introduction
%---------------------------------
\section{Introduction}
The interaction of energetic particle beams with plasmas is ubiquitous in laboratory and in astrophysical scenarios, and so are beam-plasma instabilities such as Weibel~\cite{bib:weibel_prl_1959}, current filamentation \cite{bib:fried_pf_1959, bib:silva_pop_2002} and two stream~\cite{bib:pierce_jap_1948, bib:bohm_pr_1949}. The first two instabilities, also referred as Weibel-like instabilities, are electromagnetic in nature and arise due to the anisotropy in the momentum distribution of the electrons, protons and ions. Specifically, for the current filamentation instability (CFI) the role of the velocity anisotropy is played by the counter-streaming flow of the particle beams. These instabilities generate exponentially growing magnetic fields, providing one of the possible mechanisms for generating near equipartition magnetic fields in extreme astrophysical scenarios, such as Gamma Ray Bursts (GRB)~\cite{bib:Medvedev_AJ_1999}, and are also closely associated with the formation of  relativistic Weibel mediated collisionless shocks \cite{bib:bret_pop_2013} in space~\cite{bib:gruzinov_aj_2001} and laboratory plasmas \cite{bib:fiuza_prl_2012, bib:stockem_sr_2014, bib:park_hedp_2012, bib:kugland_np_2012,bib:huntington_np_2015}. Recently, the onset of the current filamentation instability was experimentally observed in counterstreaming plasmas in high power laser experiments~\cite{bib:park_hedp_2012, bib:kugland_np_2012, bib:huntington_np_2015}. Experiments on laser wakefield acceleration have also reported the filamentation of the accelerating particle beam as it interacts with the background plasma \cite{bib:huntington_prl_2011}. These instabilities provide an efficient way of restoring the isotropy in collisionless plasmas, since the energetic particles scatter off the self generated magnetic fields by which the longitudinal momentum is transferred to the transverse momentum.  \\
The available theoretical models for CFI are restricted mainly to a purely temporal analysis \cite{bib:fried_pf_1959, bib:silva_pop_2002,bib:bret_pop_2010} and do not capture any spatial characteristics of the instabilities, which can be very relevant for finite size systems~\cite{bib:park_hedp_2012, bib:kugland_np_2012,bib:huntington_np_2015,bib:allen_prl_2012,bib:muggli_arxiv_2013,bib:sari_arxiv_2014}.\\
In this paper we obtain the relativistic spatial-temporal solutions for the unstable transverse CFI modes in cold plasmas. Our work and approach are inspired by Refs.~\cite{bib:mori_prl_1994,bib:decker_pop_1996}. A single differential equation is derived to model the instability, considering only the electron response, ignoring the finite transverse dimension effects, considering a semi-infinite plasma slab and including the effects of a beam density ramp. For a step-like Heaviside beam profile analytical solutions are obtained for physically relevant and realistic initial conditions. We further obtain the quasi-static and asymptotic behavior of the solutions. The theoretical results are compared with multidimensional particle-in-cell (PIC) simulations using OSIRIS~\cite{bib:fonseca_book}. Such spatial-temporal analysis, shown in the later part of this paper, is relevant to the jets emitted by the x-ray binaries where the velocities of the jets are in relativistic range $\sim 0.6 c$~\cite{bib:perucho_aa_2010} where spatial effects in the CFI modes are significant, or to the fireball-like beams~\cite{bib:allen_prl_2012,bib:muggli_arxiv_2013,bib:sari_arxiv_2014} interacting with the plasma.\\
%---------------------------------
%   SEC.2: Theory
%---------------------------------
\section{Theory}
\begin{figure}[htbp]
\begin{center}
\includegraphics[width=\columnwidth]{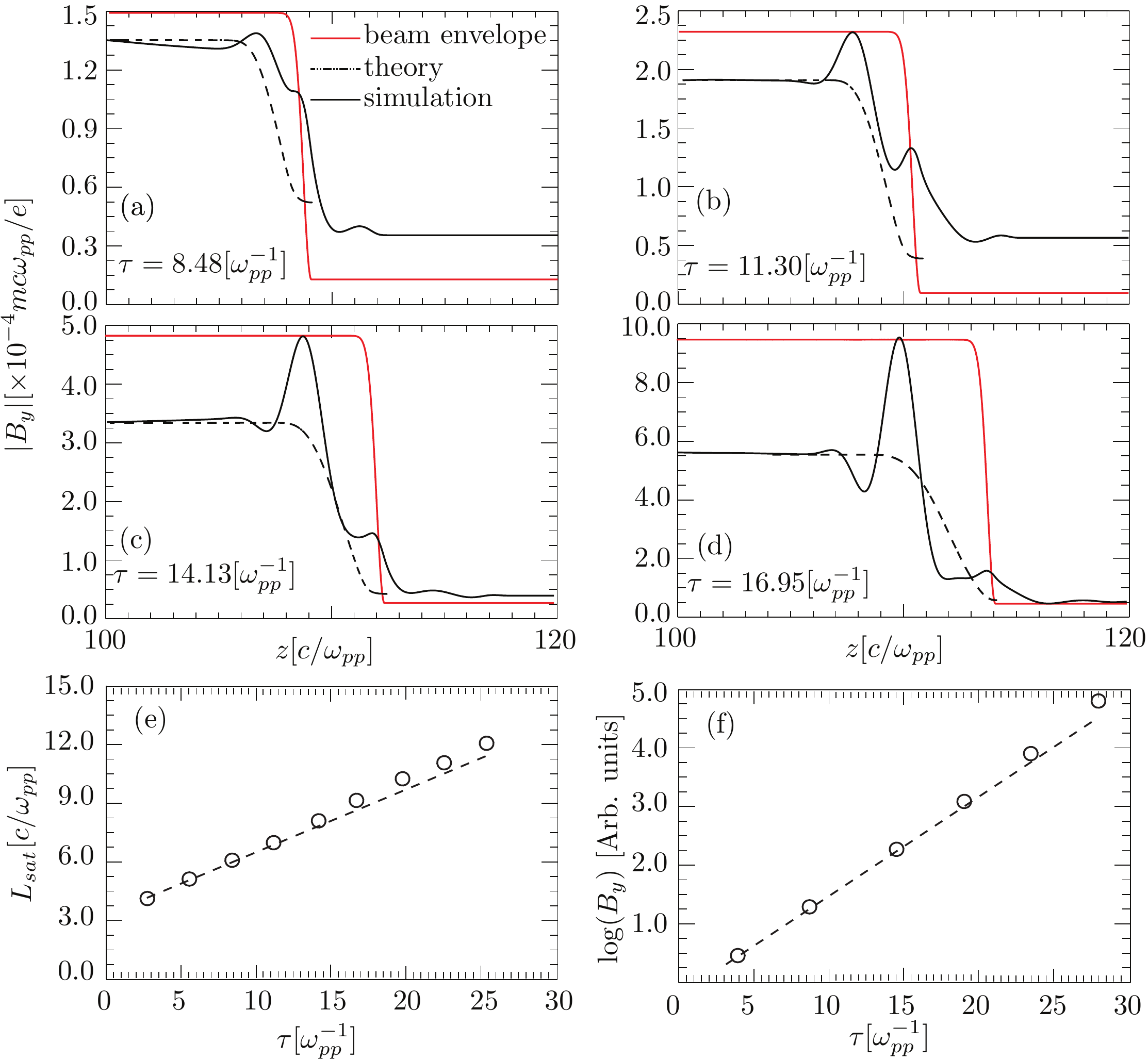}
\caption{\label{fig:sim_theo} Comparison between simulations and theory: (a), (b), (c) and (d) represent the magnetic field amplitude averaged over transverse dimension of the simulations (solid dark line), theoretical field estimates (dashed dark line) and beam profile in the simulations in arbitrary units (red/light solid line) at time $\tau_{sim}=8.48$, $11.30$, $14.13$, and $16.95 \omega_{pp}^{-1}$ respectively for $\gamma_{0b}=1.25$. To maximize the $Q$ and minimize the effect of density ramp on the spatial-temporal evolution of CFI mode we have taken $k=0.628 \omega_{pp}/c$, $n_{0b}=n_{0p}$ and $L=0.5 c/\omega_{pp}$. (e) represents temporal evolution of the $\mathrm{L_{sat}}$ ($\circ$) compared with the theoretical estimates [Eq.~(\ref{eq:eq_fil})] (dashed line: $Q=\partial_\tau \mathrm{L_{sat}}$) and (f) represents the logarithmic growth of the field in the region of purely temporal growth ($\psi \gg \psi_T$). Simulation results in $\circ$, and theoretical growth rate $\Gamma_0$ in dash line].  }
\end{center}
\end{figure}

We consider a two dimensional (2D) slab geometry, where a relativistic beam with velocity $v_{0b}\hat{z}$ and density $n_{0b}F(z,t)$, where $F(z,t)$ is the initial density profile of the beam, is propagating in a stationary plasma comprised of cold electrons and immobile ions with homogeneous plasma density $n_{0p}$. We analyze the stability of a transverse CFI mode with wavenumber $k$, and vector potential 
 $\vec{A}=A(\psi,\tau)\hat{z}~\mathrm{exp}[i k x]$, where $\psi=v_{0b} t-z$ and $\tau=t$, which satisfies the Coulomb gauge condition ${\triangledown}\cdot\vec{A}=0$  by solving the wave equation $(\triangledown^2-\partial_t^2/c^2)\vec{A}=-4 \pi \vec{J}/c$. Under the slow envelope approximation $|\partial_\psi A|\ll | k A|$, the governing equation for the vector potential of the electromagnetic wave driven by a current density $J_z$ can be written as
% We can deduce the wave equation [ $(\partial_t^2/c^2-\triangledown^2)\vec{A}=4 \pi \vec{J}/c$ ]as,
\begin{eqnarray}
\label{eq:weq_g}
\left[\frac{1}{c^2}\partial_\tau^2+\frac{2v_{0b}}{c^2}\partial_{\psi \tau}^2+k^2 \right] A = \frac{4 \pi}{c^2}J_z \mathrm{e}^{-i k x},
%\left[\frac{1}{c^2}\partial_\tau^2+\frac{2v_{0b}}{c^2}\partial_{\psi \tau}^2 -\frac{1}{\gamma_{0b}^2}
%\partial_\psi^2+k^2 \right] A = \frac{4 \pi}{c^2}J_z \mathrm{e}^{-i k x},
 \end{eqnarray}
where $\vec{J}=-e[n_{0b}F(\psi) \vec{v}_{1b}+n_{0p} \vec{v}_{1p}+n_{1b} \vec{v}_{0b}]$ is the current density 
 driving the vector potential $A\equiv A_z$, $\gamma_{0b}=1/\sqrt{1-v_{0b}^2/c^2}$ is the beam Lorentz factor, and $c$ is the velocity of light in vacuum. The suffixes $0$ and $1$ are the $0^{th}$ and $1^{st}$ order perturbed values of the plasma (p) 
 and beam (b) parameters defined as plasma electron velocity and density $\vec{v}_p=\vec{v}_{1p}$, 
 $n_p=n_{0p}+n_{1p}$, and beam electron velocity and density $\vec{v}_b=\vec{v}_{0b}+\vec{v}_{1b}$, 
 $n_b=n_{0b} F(\psi)+n_{1b}$. The chosen vector potential perturbation will generate a magnetic field $\vec{B}=\vec{\triangledown}\times \vec{A}=-ikA(\psi,\tau)\hat{y}~\mathrm{exp}[i k x]$ in the $\hat{y}$ direction. Resorting to the fluid equations of motion of a two-species-plasma (plasma electrons and beam electrons), using the continuity equation and the equation of momentum conservation for the relativistic beam and the stationary background plasma electrons, and restricting to the first order values in the weakly coupling limits by ignoring the  ($\partial_\tau+v_{0b}\partial_\psi$) term with respect to $kv_{0b}$, the perturbed quantities can be written as,
\begin{eqnarray}
\label{eq:eq_v1_n1}
\vec{v}_{1p}=-\frac{e}{m c} \vec{A}, \nonumber \\
\partial_\tau [\gamma_{0b}\vec{v}_{1b}+\gamma_{0b}^3 v_{0b}^2 v_{1bz}\hat{z}]=-\frac{e}{mc}i k v_{0b} A \mathrm{e}^{i k x} \hat{x}-\frac{e}{mc}(\partial_\tau+v_{0b}\partial_\psi)\vec{A},\nonumber  \\ 
\partial_\tau^2 n_{1b}=\frac{n_{0b} e F(\psi)}{m c^2 \gamma_{0b}} \left[ -k^2 v_{0b} c+\frac{1}{\gamma_{0b}^2}\partial_\tau(\partial_\tau+v_{0b}\partial_\psi)\right]A \mathrm{e}^{i k x}.
\end{eqnarray}
Incorporating Eq.~(\ref{eq:eq_v1_n1}) in Eq.~(\ref{eq:weq_g}) by taking the second order $\tau$ derivative of Eq.~(\ref{eq:weq_g}) and further neglecting the higher order derivatives $\partial_\tau^2, \partial_\psi\partial_\tau$, $\partial_\psi^2$  when compared with $k^2 c^2$, we obtain,
\begin{equation}
\label{eq:eq_fil}
\left[\partial_\tau^2+QF(\psi)\partial_{\psi \tau}^2-\Gamma_0^2F(\psi)\right] A=0,
\end{equation}
 where  $\Gamma_0=k v_{0b} \omega_{pb}/\sqrt{\gamma_{0b} D}$,
$D=k^2c^2+\omega_{pp}^2+\omega_{pb}^2/\gamma_{0b}^3$,
 $Q=2\omega_{pb}^2 v_{0b}/(\gamma_{0b}^3 D)$, $\omega_{pb}=\sqrt{n_{0b}e^2/(m \epsilon_0)}$ 
 and $\omega_{pp}=\sqrt{n_{0p}e^2/(m \epsilon_0)}$. Considering an infinite beam [$F(\psi)=1$] and ignoring the second term in Eq.~(\ref{eq:eq_fil}) we retrieve the well known purely temporal evolution of the system with growth rate $\Gamma_0$. 
 Interestingly, the equation obtained by Mori \textit{et al.}~\cite{bib:mori_prl_1994} (Eq. 10 in Ref.~\cite{bib:mori_prl_1994}) to analyze the spatial-temporal evolution of 
 Raman forward scattering has the same form as Eq.~(\ref{eq:eq_fil}) obtained here for the case of CFI. Equation~(\ref{eq:eq_fil}) can be solved numerically for any general beam profile; however, to obtain analytical expression, we assume $F(\psi)=H(\psi)$, where $H(\psi)= 0$ for $\psi<0$, $H(\psi)=1$ for $\psi>0$ is the Heaviside function. Respecting causality, we can impose $A=0$ for $\tau<0$, and define the double Laplace transform of  $A(\tau,\psi)$ as 
 \begin{equation}
\label{eq:lap}
A(\alpha, \beta)=\int_0^\infty d \tau \int_0^\infty d \psi A(\tau, \psi)\mathrm{exp}[-i \alpha \tau-i \beta \psi].
\end{equation}
Thus, by doing the double Laplace transformation of Eq.~(\ref{eq:eq_fil}), according to Eq.~(\ref{eq:lap}), we obtain the field expression in Laplace space  as
\begin{equation}
\label{eq:lapA}
A(\alpha, \beta)= \frac{Q A(0,0)-i \alpha \left(1+\frac{\beta}{\alpha}Q\right) A(0, \beta)-i \alpha Q 
A(\alpha,0)-\frac{\partial A}{\partial \tau} (0,\beta)}{\alpha^2+Q \alpha \beta 
+\Gamma_0^2},
\end{equation}
where $A(0,0)$, $A(0,\beta)$, $A(\alpha,0)$ and $\partial_\tau A(0,\beta)$ are the Laplace transforms of 
$A(\tau=0,\psi=0)$, $A(\tau=0,\psi)$, $A(\tau,\psi=0)$ and $\partial_\tau A(\tau=0,\psi)$ respectively, which are 
the required initial conditions. The field $A(\tau,\psi)$ can be obtained by performing inverse Laplace
 transformation of $A(\alpha, \beta)$, defined as
\begin{equation}
\label{eq:inv_lap}
A(\tau,\psi)= \frac{1}{4 \pi^2}\int_{-\infty -i \sigma_\alpha}^{\infty -i \sigma_\alpha} d \alpha 
\int_{-\infty -i \sigma_\beta}^{\infty -i \sigma_\beta} d \beta A(\alpha, \beta)\mathrm{e}^{i \alpha \tau+i \beta \psi},
\end{equation}
where $ \sigma_{(\alpha,\beta)}$ are chosen such that  the contour from $\infty -i\sigma_{(\alpha,\beta)}$ 
to $-\infty -i\sigma_{(\alpha,\beta)}$ lies below all the singularities. For the sake of simplicity we consider
 the following realistic initial conditions,
 \begin{equation}
\label{eq:con}
A(\tau,\psi=0)=A(\tau=0,\psi)=A_n,~\mathrm{and}~\partial_\tau A(\tau,\psi=0)=0,
\end{equation}
which considers that at $\tau=0$, there is an initial constant noise source throughout the beam and for $\tau>0$ the beam front ($\psi=0$) always encounters fresh plasma, and hence a constant noise source. The noise source for most instabilities are considered to be associated with the thermal fluctuations, and if thermal fluctuations have no time or space dependence, the constant noise source assumption holds correct. Longitudinally modulated or time dependent noise amplitude can be some of the forms of noise source that should be considered and the detailed analysis of the effect of different noise sources on the CFI spatial-temporal evolution will be addressed elsewhere. The above conditions yield $A(0,\beta)=A_n/(i \beta)$, $A(\alpha,0)=A_n/(i \alpha)$, $A(0,0)=A_n$ and $\partial_\tau A(0,\beta)=0$, which leads to the solution of Eq.~(\ref{eq:eq_fil}), by inverting Eq.~(\ref{eq:lapA}), as
 \begin{eqnarray}
 \label{eq:soln_exact}
 A(\tau, \psi)&=&A_n \bigg[[\mathrm{H}(\tau)-\mathrm{H}(\tau-\psi/Q)]A_n \mathrm{cosh}(\Gamma_0 \tau) 
  \nonumber \\
 &&+\mathrm{H}(\tau-\psi/Q)] \sum_{j=0}^{\infty} \left(\frac{\psi/Q}{\tau-\psi/Q}\right)^j \mathrm{I}_{2j}
  \bigg[ 2 \Gamma_0 \sqrt{\frac{\psi}{Q} \bigg(\tau-\frac{\psi}{Q}\bigg)}\bigg]\bigg],
 \end{eqnarray}
 where $\mathrm{I_j}$ is the $j^{th}$ order modified Bessel
 function of the first kind \cite{bib:abramowitz_book}. Neglecting the term $\partial_\tau^2$ in Eq.~(\ref{eq:eq_fil}) leads to the 
 quasi-static solutions, which are valid at the beam front for $\psi \ll Q \tau$, as
 \begin{eqnarray}
 \label{eq:soln_quasi}
 A(\tau, \psi)=A_n \mathrm{H}(\tau)\mathrm{H}(\psi) \mathrm{I}_{0} \bigg[ 2 \Gamma_0 \sqrt{\frac{\psi}{Q} 
 \tau}\bigg].
 \end{eqnarray}
%---------------
% Figure 1
%---------------
\begin{figure}[htbp]
\begin{center}
\includegraphics[width=\columnwidth]{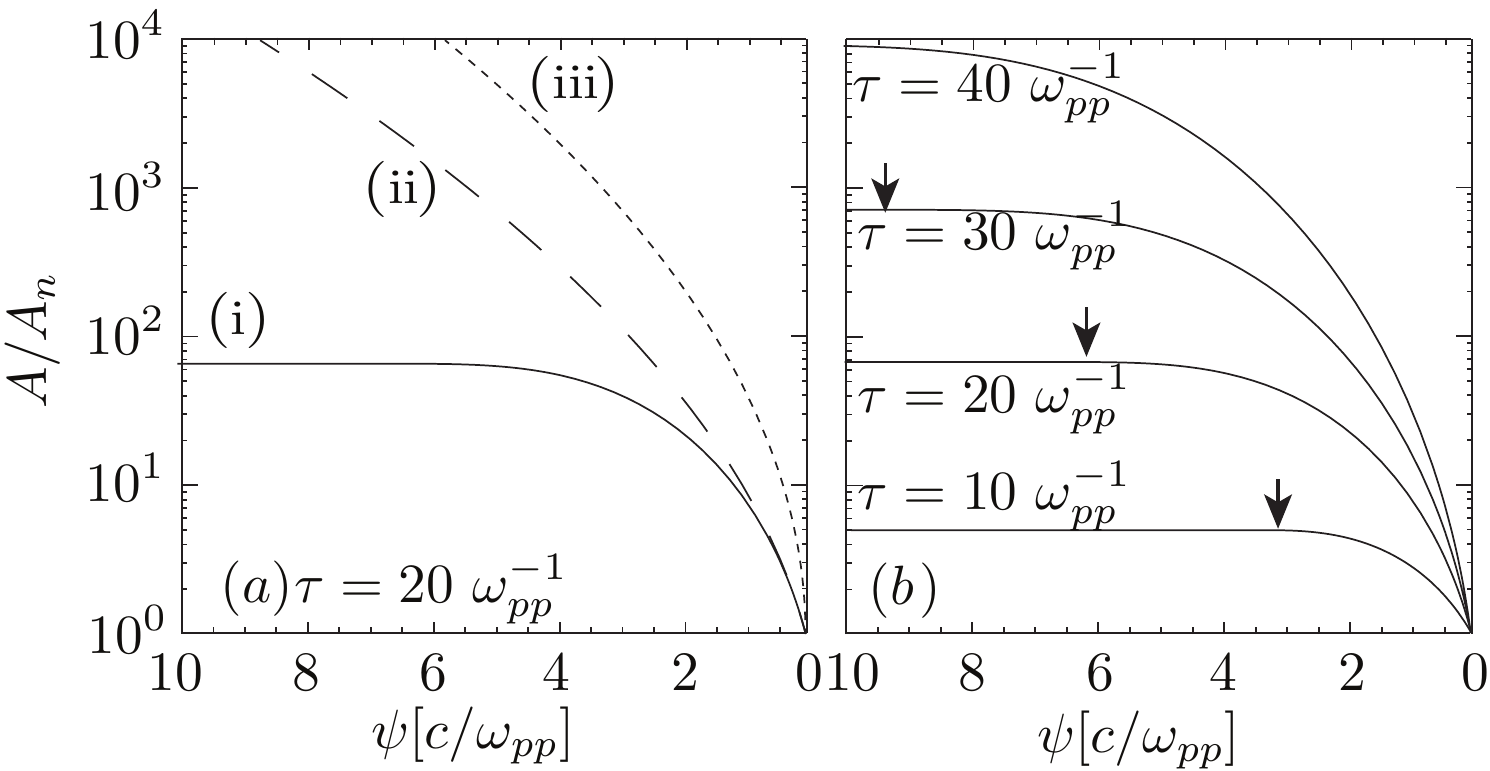}
\caption{\label{fig:theory} Evolution of the field $A/A_n$, (a) comparing (i) the exact [Eq.~(\ref{eq:soln_exact})], (ii) the quasi-static [Eq.~(\ref{eq:soln_quasi})], and (iii) the asymptotic solutions[Eq.~(\ref{eq:asymp})] of Eq.~(\ref{eq:eq_fil}) for the spatial-temporal evolution of purely transverse CFI modes at $\tau=20 \omega_{pp}^{-1}$, and (b) showing the temporal evolution of  Eq.~(\ref{eq:soln_exact}) at different times and along the beam. The beam is propagating with $\gamma_{0b}=1.25$ along the z direction in the equally dense  ($n_{0b}=n_{0p}$) plasma. The wavenumber of the CFI mode is $k=0.628 \omega_{pp}/c$ taken to get maximum $Q=0.32 c$ at $\gamma_{0b}=1.25$. The arrow pointers on the lines indicates $L_\mathrm{sat}$, which varies as $0.32c \tau$.}
\end{center}
\end{figure}
Moreover, and using the stationary phase method which gives the impulse response due to a localized initial disturbance at $\tau=0$ and $\psi=0$, the asymptotic solution for $A(\tau,\psi)$ at large $\tau$ can be written as \cite{bib:mori_prl_1994}
\begin{equation}
\label{eq:asymp}
A(\tau, \psi)=A_n\mathrm{ exp}(2\Gamma_0 \sqrt{\tau \psi/Q}).
\end{equation}
The partial differential equation governing the CFI [Eq. (\ref{eq:eq_fil})] and its exact solution [Eq. (\ref{eq:soln_exact})] are valid for $Q \tau \gg 1/k$, whereas the asymptotic solutions are valid for $Q \tau \gg 2 \Gamma_0/(Q k)$. \\
It is evident from figure~\ref{fig:theory}(a), that the quasi-static and asymptotic methods [$(ii)$ and $(iii)$ in figure~\ref{fig:theory} (a)] fail to capture the spatial saturation of the instability at the back of the beam as demonstrated by the full exact solution of Eq.~(\ref{eq:eq_fil}) [line (i) in figure~\ref{fig:theory}]. This specific characteristic is also evident in the simulation results to be discussed later in this paper. It is worth mentioning here that the asymptotic approach, used extensively  for spatial-temporal analysis of the longitudinal beam-plasma instabilities~\cite{bib:rostomian_pop_2000}, overestimates the growth and does not seem to give correct spatial characteristics for the transverse instabilities, specifically for the CFI discussed in here.\\
%---------------
% Figure 2
%---------------
\begin{figure}[htbp]
\begin{center}
\includegraphics[width=\columnwidth]{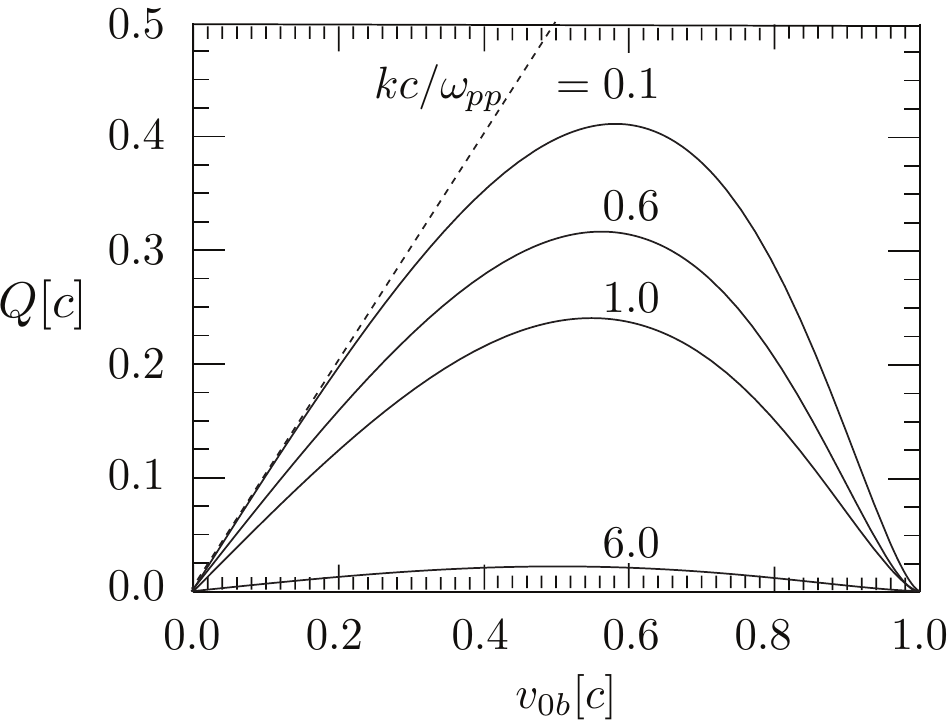}
\caption{\label{fig:Q} Dependence of cross coupling coefficient $Q$ with beam velocity $v_{0b}$ at $n_{0b}=n_{0p}$. The dashed line has a slope of $1$, plotting $Q=v_{0b}$. We note that large $Q$ implies a larger region behind the beam front where the spatial-temporal behavior is significant.}
\end{center}
\end{figure}
At the beam front, for $\psi \ll Q \tau$ the quasi-static solutions given by Eq.~(\ref{eq:soln_quasi}) match well with the exact solutions described by Eq.~(\ref{eq:soln_exact}). The mildly relativistic ($\gamma_{0b}=1.25$) solutions for $A(\tau, \psi)$ [Eq.~(\ref{eq:soln_exact})], presented in figure~\ref{fig:theory} (b) with respect to $\psi$ for different  times $\tau$, indicate that the filaments grow spatially from the beam front ($\psi=0$) to the back until the transition point $\psi_T=Q \tau$. After the transition point the instability grows in a purely temporally fashion. We define the beam length over which the instability grows spatially as $\mathrm{L_{sat}}=Q\tau$ [identified as vertical arrows in each line in figure~\ref{fig:theory} (b)]. Beyond this length the instability grows with spatially constant temporal growth rate $\Gamma_0$.\\
As observed from the previous discussions, and from Eqs.~(\ref{eq:eq_fil}), (\ref{eq:soln_exact}), (\ref{eq:soln_quasi}) and (\ref{eq:asymp}), the spatial-temporal behavior depends on the cross coupling coefficient $Q$. To address this, in figure~\ref{fig:Q} we analyze the dependence of $Q$ on the beam velocity $v_{0b}$ for different CFI wavenumbers $k$ at $n_{0b}=n_{0p}$.  For $kc \gg \omega_{pp},~\omega_{pb}$, the maximum value of $Q$ is achieved for $v_{0b}=0.5c ~ (\gamma_{0b}=1.15)$, and varies as $Q_{max}\simeq 0.65 n_{0b}/(n_{0p}k^2c^2)$. For $kc \ll \omega_{pp}=\omega_{pb}$, $Q_{max}\simeq 0.4$ at $v_{0b}=0.6 c~ (\gamma_{0b}=1.25)$. In the non-relativistic scenario $\gamma_{0b}\simeq 1$ and $kc \ll \omega_{pp}=\omega_{pb}$, $Q=v_{0b}$, which is shown as a dashed line in figure~\ref{fig:Q}. At higher $\gamma_{0b} \gg 1$, $k>>\omega_{pp}/c$ or $n_{0b} \ll n_{0p}$, $Q$ tends to $0$, and the instability acquires a purely temporal behavior.\\
We have also considered a beam profile with $F(\psi)=1-\mathrm{exp}[-\psi^2/L^2]$, of direct relevance for the comparison with simulations. For such beam configurations the numerical solution of Eq. (\ref{eq:eq_fil}) gives the same
spatial-temporal behavior predicted by Eq. (\ref{eq:soln_exact}) but with an enhanced saturation length $L_\mathrm{sat}\simeq Q \tau+L$. The results are compared in figure~\ref{fig:sim_theo} which will be discussed in connection with the simulations performed in the next section. For $F(\psi)=1-\mathrm{exp}[-\psi^2/L^2]$, the beam density profile, and hence the effective temporal growth rate $ \propto \Gamma_0 \sqrt{F}$, reaches the maximum density growth rate on the spatial scale length $L$. Thus, in presence of a density ramp the spatial evolution of CFI can be attributed both to the beam density spatial profile, and to the cross coupling term.  If $L \gg Q \tau$, we may ignore the contribution from the cross coupling term ($Q \partial_{\psi \tau}^2$) in Eq.~(\ref{eq:eq_fil}), resulting in a field varying as $A=A_0\mathrm{cosh}[\Gamma_0 \tau \sqrt{1-\mathrm{e}^{-\psi^2/L^2}}]$, thus determining an extra condition for the relevance of the spatial-temporal effects of the current filamentation instability.
%---------------
% Figure 3
%---------------
%---------------------------------
%   SEC.3: Simulation
%---------------------------------
\section{Simulations}
In order to confirm and to explore the theoretical results presented above, we have performed 2D PIC simulations using OSIRIS~\cite{bib:fonseca_book}. We consider a scenario where a relativistic beam is propagating through a cold plasma, 
where the beam and the plasma are both comprised of an electron-proton neutral plasma (the temperature of the background plasma is set to zero). The simulation box, with dimensions $250\times 100 (c/\omega_{pp})^2$, is divided into $12500 \times5000$ cells with $3 \times 3$ 
particles per cell per species. The beam has a gaussian density ramp at the front,  $F(\psi)=1-\mathrm{exp}[-\psi^2/L^2]$,
 where $L$ is the length of the gaussian ramp at the beam front. When $L\to 0$ this profile mimics a sharp rising beam front with a Heaviside function profile which was considered to obtain the analytical solutions [Eq.~(\ref{eq:soln_exact})]. We seed the instability, in order to analyze a single CFI mode with wavenumber $k_\mathrm{seed}$,  with a small magnetic field perturbation of the form $B_y(\tau=0)= \delta B_0 \mathrm{cos}(k_\mathrm{seed} x)$, where $\delta B_0=5\times 10^{-5} mc\omega_{pp}/e$. \\
The comparisons between the CFI magnetic field $B_y$ evolution predicted by the theory [numerical solutions of Eq.~(\ref{eq:eq_fil})] and the fields observed in the simulations, plotted in figure~\ref{fig:sim_theo}, show that the solutions given by Eq.~(\ref{eq:soln_exact}) are the  most suitable model, among the three models discussed here to predict the spatial-temporal growth of the CFI along the beam, as expected. In the simulations the instability starts to grow after a relaxation time ($\tau_\mathrm{relax} \approx 1.63 \omega_{pp}^{1}$ for this particular simulation) necessary for the self-consistent electromagnetic fields and the electromagnetic noise to adjust to the initial flow condition. Thus, for the comparison with the theory, the time is re-normalized to $\tau=\tau_\mathrm{sim}-\tau_\mathrm{relax}$, where $\tau_\mathrm{sim}$ is the simulation time. One can observe in figure~\ref{fig:sim_theo} that the theoretical estimate for the CFI magnetic field, given by Eq.~(\ref{eq:soln_exact}), matches well with the magnetic field profile observed in the simulations. We analyze the variation of saturation length $\mathrm{L_{sat}}$ with time $\tau$ in figure \ref{fig:sim_theo} (e) obtained from the simulations. The rate at which $L_\mathrm{sat}$ increases with time $\tau$ is $Q=0.32 c$, which is equal to the theoretical value of $Q=\partial_\tau \mathrm{L_{sat}}$, as predicted by our model. Beyond the beam length $\mathrm{L_{sat}}$, the magnetic field amplitude is spatially constant and grows temporally with growth rate $\Gamma_0$, as predicted by the theory. \\
%---------------
% Figure 4
%---------------
\begin{figure}[htbp]
\begin{center}
\includegraphics[width=\columnwidth]{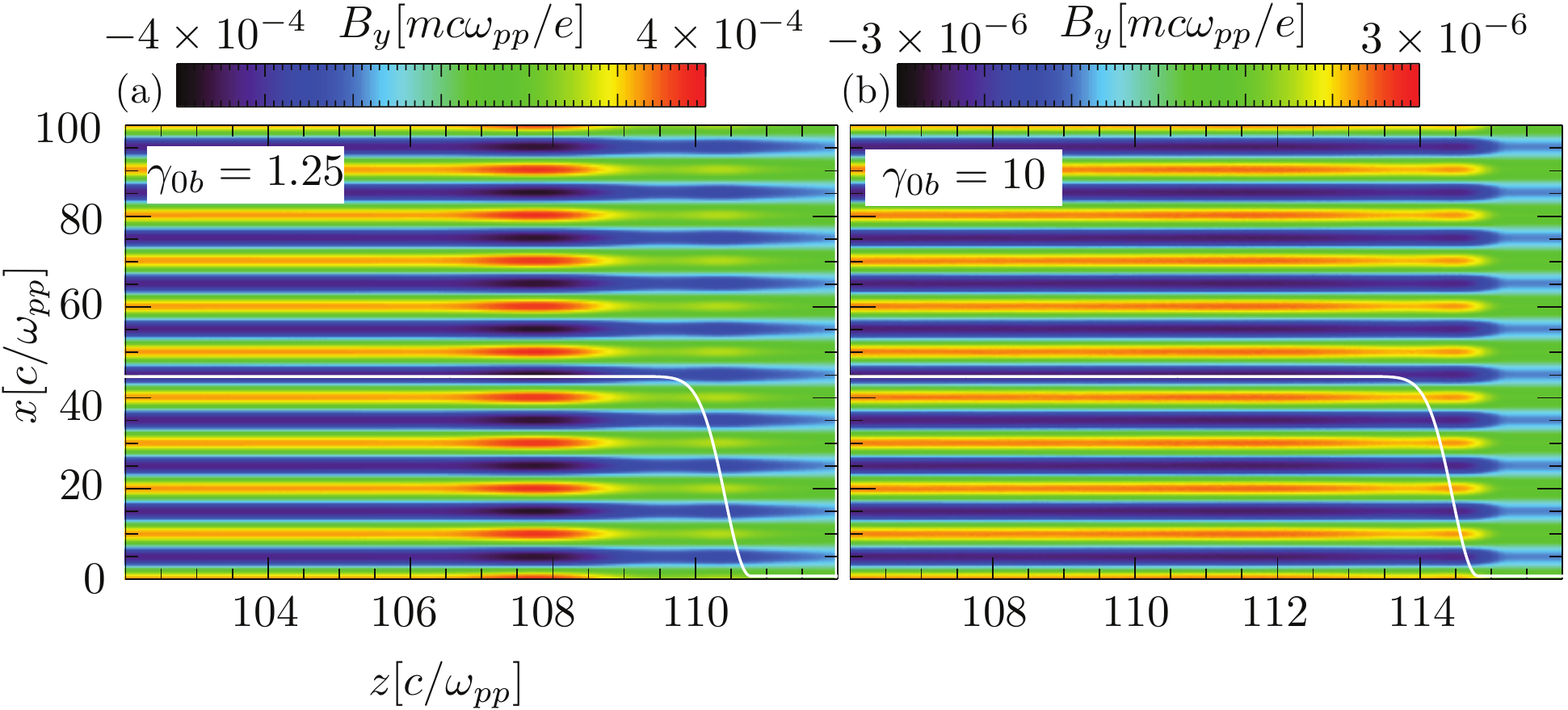}
\caption{\label{fig:snap} Effect of $\gamma_{0b}$ on spatial evolution of the current filamentation instability: Snap shots of magnetic field $B_y$ at time $=11.30 \omega_{pp}^{-1}$ for (a) $\gamma_{0b}=1.25$ and (b)$\gamma_{0b}=10$, demonstrating that at high Lorentz factor the spatial properties of the instabilities are negligible. Other parameters are: $k_\mathrm{seed}=0.628 \omega_{pp}/c$, $n_{0b}=n_{0p}$ and $L=0.5 c/\omega_{pp}$. The white line shows the beam profile in the simulations. After time $\tau > 11.30 \omega_{pp}^{-1}$ longitudinal modulations on the CFI becomes significantly strong to visualize the spatial saturation of the fields. }
\end{center}
\end{figure}
 %---------------
% Figure 5
%---------------
\begin{figure}[htbp]
\begin{center}
\includegraphics[width=\columnwidth]{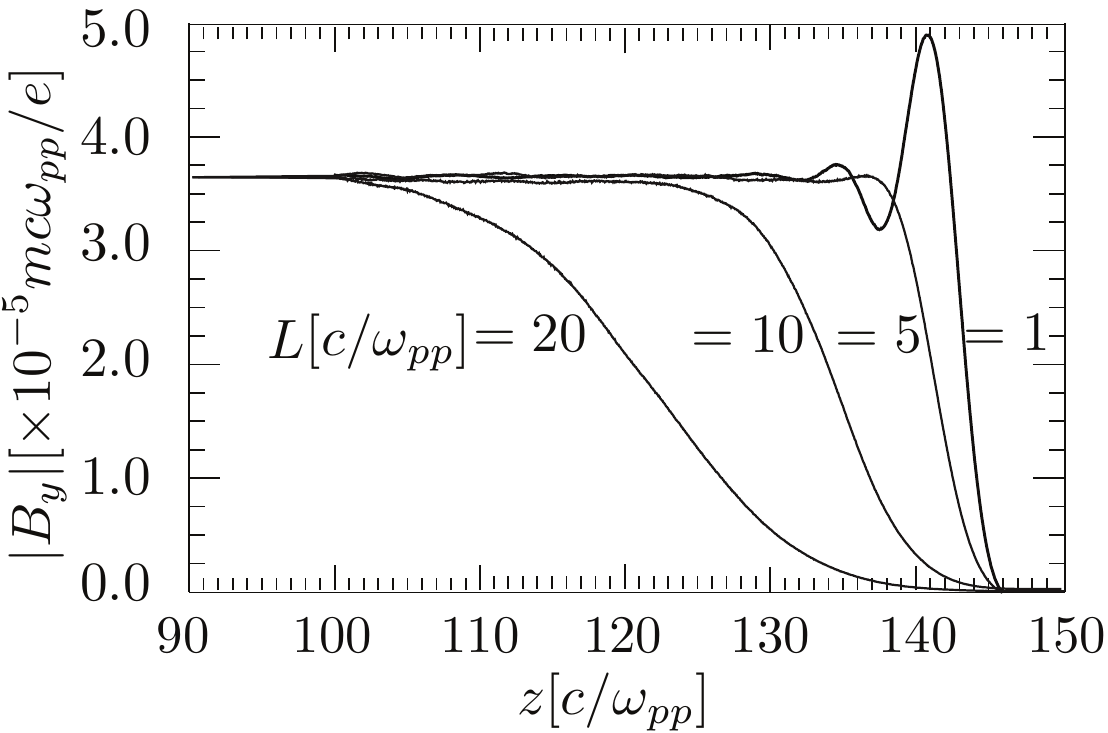}
\caption{\label{fig:By_L} The current filamentation fields growing along the beam at time $\tau_{sim}=45 \omega_{pp}^{-1}$ in the simulations performed with varying the size of density ramp $L$ for $\gamma_{0b}=10$ and $k=6.0 \omega_{pp}/c$. The longitudinal mode responsible for the modulations at the front of the beam are suppressed for longer density ramps due to the reduced noise level for excitation of the longitudinal modes.}
\end{center}
\end{figure}
The longitudinal modulations, with wavelength $\sim \lambda_p$, seen in the simulations of figure \ref{fig:sim_theo} are due to the growth of the longitudinal instability seeded by the sharp rising beam density profile at the front. In the simulations, the longitudinal electric field modulation is observed, but confined only in the front portion of the beam. The remainder of the beam does not show any sign of the longitudinal electrostatic instability. We attribute this to the fact that, in a similar way as for the CFI, the longitudinal instabilities also have a spatial-temporal nature \cite{bib:rostomian_pop_2000}. This also demonstrates that a full understanding of this scenario requires the combined analysis of CFI and longitudinal electrostatic instabilities.\\
The magnetic field snapshots in the $x-z$ plane, shown in figure~\ref{fig:snap} indicate a sharper rise in the magnetic field at the beam front for higher $\gamma_{0b}$ [figure~\ref{fig:snap} (b)] as compared to lower $\gamma_{0b}$ [figure~\ref{fig:snap} (a)], which further validates the theory, since at high $\gamma_{0b}$ the cross coupling term $Q$, and thus the saturation length $\mathrm{L_{sat}}$, decreases as $\sim1/\gamma_{0b}^3$ for a given time $\tau$. Since $\Gamma_0 \propto 1/\sqrt{\gamma_{0b}}$, the field amplitude at the back of the beam in figure~\ref{fig:snap} (b) (high $\gamma_{0b}$) is weaker as compared to the field in figure~\ref{fig:snap} (a) ( low $\gamma_{0b}$), also as predicted by the theory.\\
As observed in figure~\ref{fig:By_L}, where the transversely averaged $B_y$ is compared for various density ramps, on increasing the ramp size L, and thus reducing the initial seed for the longitudinal modes, the simulation results show that these longitudinal modulations on the purely transverse CFI modes (at the wavelength $\approx \lambda_p$) can be suppressed. As the time progresses (not shown in the paper) the longitudinal modes as well as other faster growing CFI modes start to play an important role and their interplay in the nonlinear stage becomes significant.\\

%---------------------------------
%   SEC.4: Conclusion
%---------------------------------
\section{Discussion and Conclusions}
 To summarize, in this paper we have derived a single differential equation modeling the spatial-temporal evolution of the purely transverse current filamentation instability. For relevant initial conditions exact analytical solutions have been obtained and compared with the analytical solutions under the quasi-static and asymptotic approximations. The validity of the model was demonstrated by comparing it with 2D PIC simulations in OSIRIS~\cite{bib:fonseca_book}. In a setup of a cold relativistic beam propagating in a uniform cold plasma the instability grows from the beam front to the back, acquires maximum value at the critical beam length $\mathrm{L_{sat}}=Q \tau$ at given time $\tau$ and then grows in a purely temporal manner for the rest of the beam length. \\
For relativistic fireball electron-positron beams \cite{bib:muggli_arxiv_2013, bib:sari_arxiv_2014} undergoing current filamentation in an electron-ion plasma, the cross coupling coefficient $Q$ is enhanced by a factor of 2 ($Q\to 2Q$) and the purely temporal growth rate is enhanced by a factor of $\sqrt{2}$ ($\Gamma_0 \to \sqrt{2} \Gamma_0$) due to the contribution from the current driven by the velocity and density perturbations of the positrons in the beam. However, in the relativistic regime this enhancement is not sufficient to balance the $1/\gamma_{0b}^3$ dependence of $Q$ on the beam Lorentz factor $\gamma_{0b}$. As a result for an ultra-relativistic $29$ $\mathrm{GeV}$ fireball beam \cite{bib:muggli_arxiv_2013} with $n_{0b}=n_{0p}$, $Q \simeq 2 \times 10^{-14}c$, and thus the spatial evolution to the CFI can be attributed only to the density gradient scale length. For the recent experiments with a $60 \mathrm{MeV}$ electron beam \cite{bib:allen_prl_2012}, $Q \simeq 1.2 \times 10^{-6}c$, thus again suggests only purely temporal growth of the CFI is present along the beam.  However, in the case of moderately relativistic fireball beams $Q$ can be significantly enhanced. For instance, in the case of Sarri \textit{et al.} \cite{bib:sari_arxiv_2014} with $\gamma_{0b}=15$, $n_{0b}=10 n_{0p}$ and considering a density ramp of $L=0.22 c/\omega_{pp}$, the cross coupling coefficient can be $Q \approx 0.01 c$, which suggests that in the linear regime of the CFI, for $\tau \gg 220 \omega_{pp}^{-1}$ the CFI spatially grows beyond the density ramp size $L$  and spatially saturates with  $\mathrm{L_{sat}}\simeq Q \tau \gg L$. \\
Based on our analysis we further observe that the spatial-temporal nature of the instability also has an effect for finite beam-plasma interaction time $\tau_\mathrm{int}$ and beam size $L_\mathrm{beam}$. In fact, depending on the relation between these parameters, \textit{i.e.} either $ L_\mathrm{beam}< Q \tau_\mathrm{int}$ or $ > Q \tau_\mathrm{int}$, a weaker or stronger filamentation of the beam can be expected which will further affect the nonlinear growth of the instability. Moreover, the study of spatial-temporal evolution of the beam-plasma instabilities can also lead to a better understanding and characterization of the Weibel mediated collisionless shock formation process in laboratory and astrophysical plasmas.\\

\textbf{Acknowledgement}\\

This work was supported by the European Research Council (ERC-2010-AdG grant 267841), Institute of Basic Science Korea (Project code: IBS-R012-D1) and Korea Institute of Science and Technology Information(Project code: KSC-2014-C1-049). We also acknowledge PRACE for providing access to SuperMUC based in Germany at the Leibniz research center.\\ 
\

\textbf{References}\\

\end{document}